\documentclass[aip,amsmath,amssymb,reprint]{revtex4-1}

\usepackage{graphicx}% Include figure files
\usepackage{dcolumn}% Align table columns on decimal point
\usepackage{bm}% bold math
\usepackage{chemformula}
\usepackage{tabularx}
\usepackage[utf8]{inputenc}
\usepackage[T1]{fontenc}
\usepackage{mathptmx}
\usepackage{etoolbox}
\usepackage{booktabs}
\usepackage[colorlinks=true,
            linkcolor=blue,
            citecolor=blue,
            urlcolor=blue,
            pdfborder={0 0 0}]{hyperref}
\usepackage{float}
\usepackage{array}
\usepackage{xr}
\def\@email#1#2{
 \endgroup
 \patchcmd{\titleblock@produce}
  {\frontmatter@RRAPformat}
 {\frontmatter@RRAPformat{\produce@RRAP{*#1\href{mailto:#2}{#2}}}\frontmatter@RRAPformat}
 {}{}
}
\makeatother
\begin{document}

\preprint{AIP/123-QED}

\title{How Accurate Are DFT Forces? Unexpectedly Large Uncertainties in Molecular Datasets}

\author{Domantas Kuryla}
\affiliation{Yusuf Hamied Department of Chemistry, University of Cambridge, Lensfield Road, Cambridge, UK}
\affiliation{Engineering Laboratory, University of Cambridge, Trumpington St and JJ Thomson Ave, Cambridge, UK}

\author{Fabian Berger}
\email{fb593@cam.ac.uk}
\affiliation{Yusuf Hamied Department of Chemistry, University of Cambridge, Lensfield Road, Cambridge, UK}%

\author{G\'abor Cs\'anyi}
\email{gc121@cam.ac.uk}
\affiliation{Engineering Laboratory, University of Cambridge, Trumpington St and JJ Thomson Ave, Cambridge, UK}

\author{Angelos Michaelides}
\email{am452@cam.ac.uk}
\affiliation{Yusuf Hamied Department of Chemistry, University of Cambridge, Lensfield Road, Cambridge, UK}

\begin{abstract}
Training of general-purpose machine learning interatomic potentials (MLIPs) relies on large datasets with properties usually computed with density functional theory (DFT). A pre-requisite for accurate MLIPs is that the DFT data are well converged to minimize numerical errors. A possible symptom of errors in DFT force components is nonzero net force. Here, we consider net forces in datasets including SPICE, Transition1x, ANI-1x, ANI-1xbb, AIMNet2, QCML, and OMol25. Several of these datasets suffer from significant nonzero DFT net forces. We also quantify individual force component errors by comparison to recomputed forces using more reliable DFT settings at the same level of theory, and we find significant discrepancies in force components averaging from 1.7~meV/\AA\ in the SPICE dataset to 33.2~meV/\AA\ in the ANI-1x dataset. These findings underscore the importance of well converged DFT data as increasingly accurate MLIP architectures become available.
\end{abstract}
 
\maketitle

The development of machine learning interatomic potentials (MLIPs) has been accelerated by the availability of curated molecular datasets. Over the last decade, successively larger and more chemically complex datasets have been produced \cite{ramakrishnan_quantum_2014, md_17, smith_ani-1ccx_2020, devereux_extending_2020, christensen_orbnet_2021, schreiner_transition1x_2022, eastman_spice_2023,eastman_nutmeg_2024,  
zhang_ani-1xbb_2025, anstine_aimnet2_2025, ganscha_qcml_2025, levine2025openmolecules2025omol25}. The developers of these datasets have built valuable foundations for general-purpose MLIPs that give access to fast and stable molecular simulations \cite{kovacs_mace-off_2025, 
anstine_aimnet2_2025, moore_computing_2024}.

For MLIP training, the structures are labeled with their energies and forces computed usually with density functional theory (DFT) \cite{hohenberg-kohn, kohn-sham}. The quality of DFT forces is important for the accuracy of MLIPs. Unconverged electron densities as well as numerical errors due to approximate DFT setups can degrade the accuracy of the forces used for training and testing. While errors in forces potentially affect the quality of the MLIP, they definitely affect the test error, i.e. our understanding of how accurate the MLIP is. Considering that general-purpose MLIP force mean absolute errors and root mean square errors (RMSE) are on the order of tens of meV/\AA\ and sometimes down to 10~meV/\AA \cite{batatia_foundation_2024, kovacs_mace-off_2025, levine2025openmolecules2025omol25}, errors in DFT forces should be much smaller, ideally 1~meV/\AA\ and less to not affect the MLIP quality.

A clear indicator of numerical errors in the forces is the nonzero net force. The net force is obtained by taking the sum of the force components on all the atoms for each Cartesian direction, and should be zero in the absence of external fields. However, errors in individual DFT force components often do not cancel, resulting in nonzero net forces.
In this Communication, we show that the ANI-1x, Transition1x, AIMNet2, and SPICE datasets have unexpectedly large net forces, indicating suboptimal DFT settings. We then quantify the actual errors in forces by comparing to forces obtained using well converged DFT settings.

\begin{table}[htp]
\begin{tabularx}{\columnwidth}{p{0.20\columnwidth} p{0.15\columnwidth} p{0.32\columnwidth} p{0.252\columnwidth}} 
 \toprule
 Dataset & Size & Level of theory & DFT code \\ [0.5ex] 
 \toprule
 ANI-1xbb \cite{zhang_ani-1xbb_2025} & 13.1 M & B97-3c \cite{brandenburg_b97-3c_2018} & ORCA 4 \cite{neese_orca_2020} \\
 \midrule
 QCML \cite{ganscha_qcml_2025} & 33.5 M & PBE0 \cite{pbe0} & FHI-aims \cite{fhi-aims} \\
 \midrule
 ANI-1x \cite{smith_ani-1ccx_2020} & $^{(1)}$5.0 M\newline$^{(2)}$4.6 M & $\omega$B97x \cite{wb97x}\newline $^{(1)}$6-31G* \cite{ditchfield_self-consistent_1971, hehre_selfconsistent_1972, hariharan_influence_1973},\newline$^{(2)}$def2-TZVPP \cite{karlsruhe1} & $^{(1)}$Gaussian 09\cite{gaussian09}\newline $^{(2)}$ORCA 4 \\
 \midrule
 AIMNet2 \cite{anstine_aimnet2_2025} & 20.1 M & $\omega$B97M-D3(BJ) \cite{mardirossian__2016, grimme_consistent_2010, grimme_effect_2011}\newline def2-TZVPP & ORCA 5\\
 \midrule
 Transition1x \cite{schreiner_transition1x_2022} & 9.6 M & $\omega$B97x\newline 6–31G(d) \cite{ditchfield_self-consistent_1971, hehre_selfconsistent_1972, hariharan_influence_1973} & ORCA 5.0.2 \\
 \midrule
 SPICE \cite{eastman_spice_2023, eastman_nutmeg_2024} & 2.0 M & $\omega$B97M-D3(BJ)\newline def2-TZVPPD \cite{rappoport_property-optimized_2010} & Psi4 \cite{smith_p_2020}\\
 \midrule
 OMol25 \cite{levine2025openmolecules2025omol25} & 100 M & $\omega$B97M-V \cite{mardirossian__2016} \newline def2-TZVPD \cite{rappoport_property-optimized_2010} & ORCA 6.0.0 \\ [1ex] 
 \bottomrule
\end{tabularx}
\caption{Summary of datasets considered in this work, including their sizes in terms of  number of configurations, exchange-correlation functionals, basis sets, and DFT codes used.}
\label{table:1}
\end{table}

\begin{figure*}[ht!]
    \includegraphics[width=\textwidth]{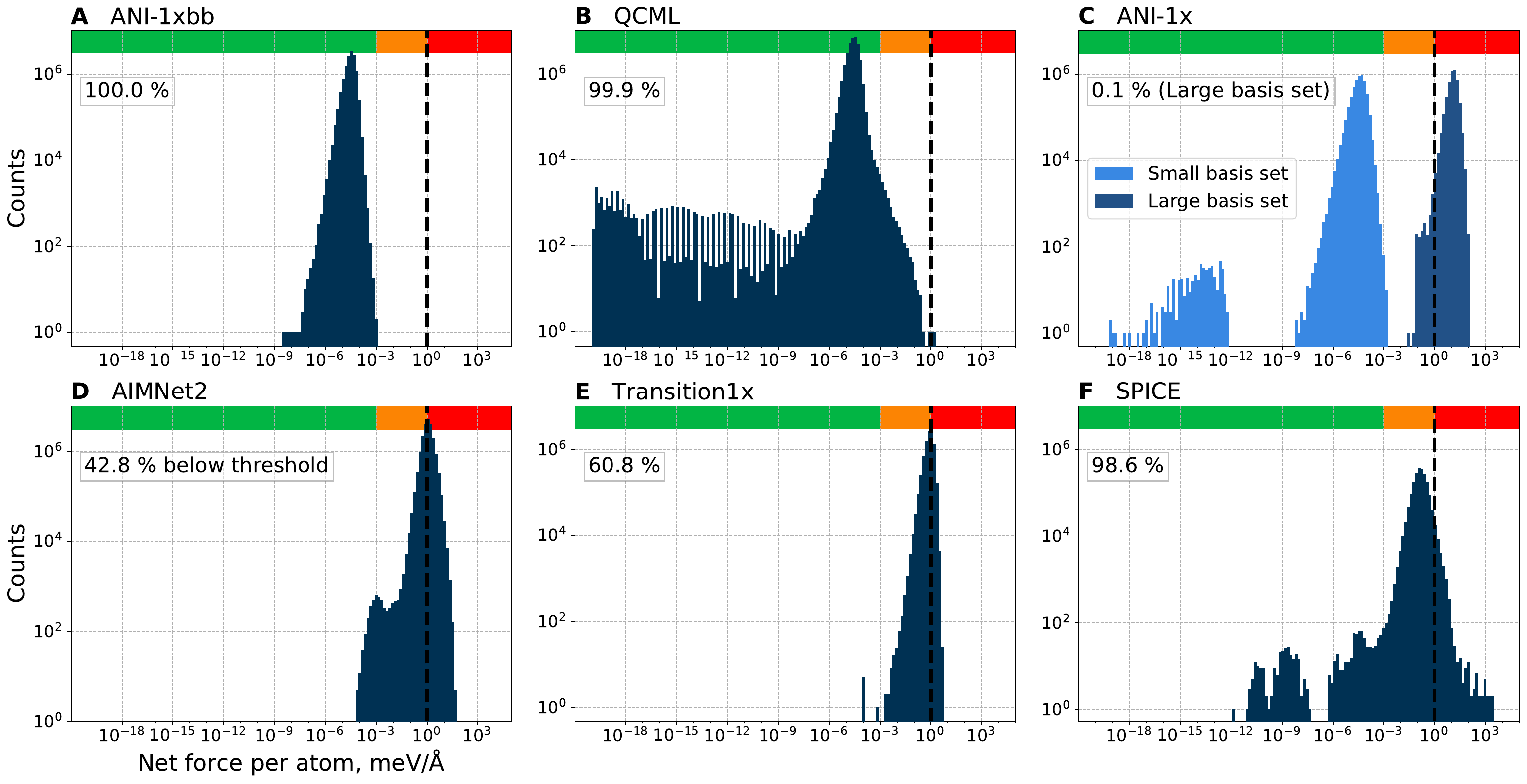}
    \caption{Distributions of net force per atom in the ANI-1xbb, QCML, ANI-1x, AIMNet2, Transition1x, and SPICE datasets. The 1~meV/\AA/atom threshold is indicated by a vertical dashed line in each plot. On the top left of each of panels \textbf{A}-\textbf{F}, the fraction of the dataset with net forces below the threshold is indicated. The red bar at the top of each panel highlights net forces above 1~meV/\AA/atom which indicate significant errors in individual force components.  In the region between $10^{-3}$ and 1~meV/\AA, indicated with an amber bar, significant DFT force errors often appear but do not result in large net forces. Negligible net forces below $10^{-3}$~meV/\AA\ are in the region indicated with the green bar.}
    \label{fig:total_forces}
\end{figure*}

The focus of this work is on molecular datasets with available DFT forces, as listed in \autoref{table:1}, specifically, the ANI-1xbb \cite{zhang_ani-1xbb_2025}, QCML \cite{ganscha_qcml_2025}, ANI-1x \cite{smith_ani-1ccx_2020}(small basis set 6-31G* and large basis set def2-TZVPP subsets, the latter used to train AIMNet \cite{zubatyuk_accurate_2019}), AIMNet2 \cite{anstine_aimnet2_2025}, Transition1x \cite{schreiner_transition1x_2022}, SPICE \cite{eastman_spice_2023, eastman_nutmeg_2024}, and OMol25 \cite{levine2025openmolecules2025omol25} datasets. We note that the AIMNet2 dataset contains data from older ANI-2x \cite{devereux_extending_2020} and OrbNet Denali \cite{christensen_orbnet_2021} datasets computed with smaller basis sets.

\autoref{fig:total_forces} shows the distribution of net force magnitudes divided by the number of atoms in each system for each dataset. For OMol25, we do not show the net force distribution, as we find that the net forces are exactly zero within numerical precision. From recomputed results discussed later, we conclude that if the net force is above 1~meV/\AA/atom for a given structure, then the RMSE in DFT force components tends to be above 1~meV/\AA. In \autoref{fig:total_forces}, we refer to the 1~meV/\AA/atom net force boundary as the threshold. In the region above this threshold, the net force indicates the presence of significant force errors, and we have shown this region in red. Below this threshold, significant force errors may still occur, but they do not manifest as large net forces as a result of error cancellation. In particular, many configurations with net forces in the region of 0.001 to 1~meV/\AA/atom still show force errors of \textgreater 1~meV/\AA, and we have indicated this region in amber. Net forces below 0.001~meV/\AA/atom are negligible, and we highlight this region in green. The datasets clearly differentiate into two groups: those with negligible net forces and those with net forces exceeding 1~meV/\AA/atom. The ANI-1xbb, QCML, and the small basis set ANI-1x datasets fall in the first category, with most or all net forces in the green region, while in the QCML dataset only a small fraction of structures have net forces in the amber region. In the other datasets, the net forces are much greater. In the large basis set part of ANI-1x, only 0.1\% of the configurations have net forces below the threshold. Notably, the large net forces occur in the data computed with a larger basis set, intended for better quality MLIP training. In AIMNet2, Transition1x, and SPICE, respectively 42.8\%, 60.8\%, and 98.6\%, of the data are below the 1~meV/\AA/atom threshold with most of it in the intermediate amber region.  The SPICE dataset also consists of individual subsets, whose net force distributions we show in section S2 of the supplementary material.

The net force acting on a system indicates numerical errors in the underlying DFT calculation. However, the net force can only be used to derive a rough estimate of the errors in the individual force components because these errors may cancel out to different degrees for different systems. To quantify the errors in individual force components arising from suboptimal DFT settings, we compare the reported forces against tightly converged reference forces computed with the same DFT functional and basis set. To compare the forces contained in the datasets with our more accurate forces, we take random samples of 1000 configurations each from the ANI-1x (large basis set), Transition1x, AIMNet2, and SPICE datasets and recompute their DFT forces using the original functional and basis set. 

To identify a set of accurate computational settings, we first identify the sources of errors. We find that, in the Transition1x, ANI-1x (large basis set), and AIMNet2 datasets computed with older ORCA 4 \cite{neese_orca_2020} and ORCA 5 \cite{neese_software_2022} versions, the nonzero net forces are eliminated by disabling the use of the RIJCOSX approximation \cite{ rijcosx}. This is an approximation used to accelerate the evaluation of Coulomb and exact exchange integrals. In the more recent ORCA~6.0.1 version \cite{neese_software_2025}, the RIJCOSX approximation no longer causes problematic net forces and results in negligible net forces on the order of $10^{-5}$ meV/\AA/atom. However, we still observe discrepancies in individual force components of up to several meV/\AA\ between forces computed with and without RIJCOSX. In addition, we explore different DFT grid settings and determine that the tightest grids provided by the \texttt{DEFGRID3} keyword are required, as less tight grids result in significant force deviations. Notably, ORCA~6 with \texttt{DEFGRID3} and RIJCOSX is used in the production of the OMOL25 dataset. The SPICE data were computed using the Psi4 software \cite{smith_p_2020}, which we compare to our recomputed ORCA forces. We also explore how to minimize the Psi4 net forces. We are able to reduce the net forces on randomly chosen SPICE systems to below 0.01 meV/\AA/atom by using denser spherical and radial DFT grids, described in section S1 of the supplementary material. We note that in ORCA~6.0.1, the D3 dispersion forces do not agree with the Psi4 D3 dispersion forces which are in agreement with the forces from the standalone D3 package \cite{grimme_consistent_2010, grimme_effect_2011}. For this reason, we compare DFT forces without the D3 dispersion correction for SPICE configurations. Based on this exploration of DFT settings, we choose to use ORCA~6.0.1 using the \texttt{DEFGRID3} grid setting and with the RIJCOSX approximation disabled to obtain accurate reference forces.

\begin{figure}[ht!]
\includegraphics[width=\columnwidth]{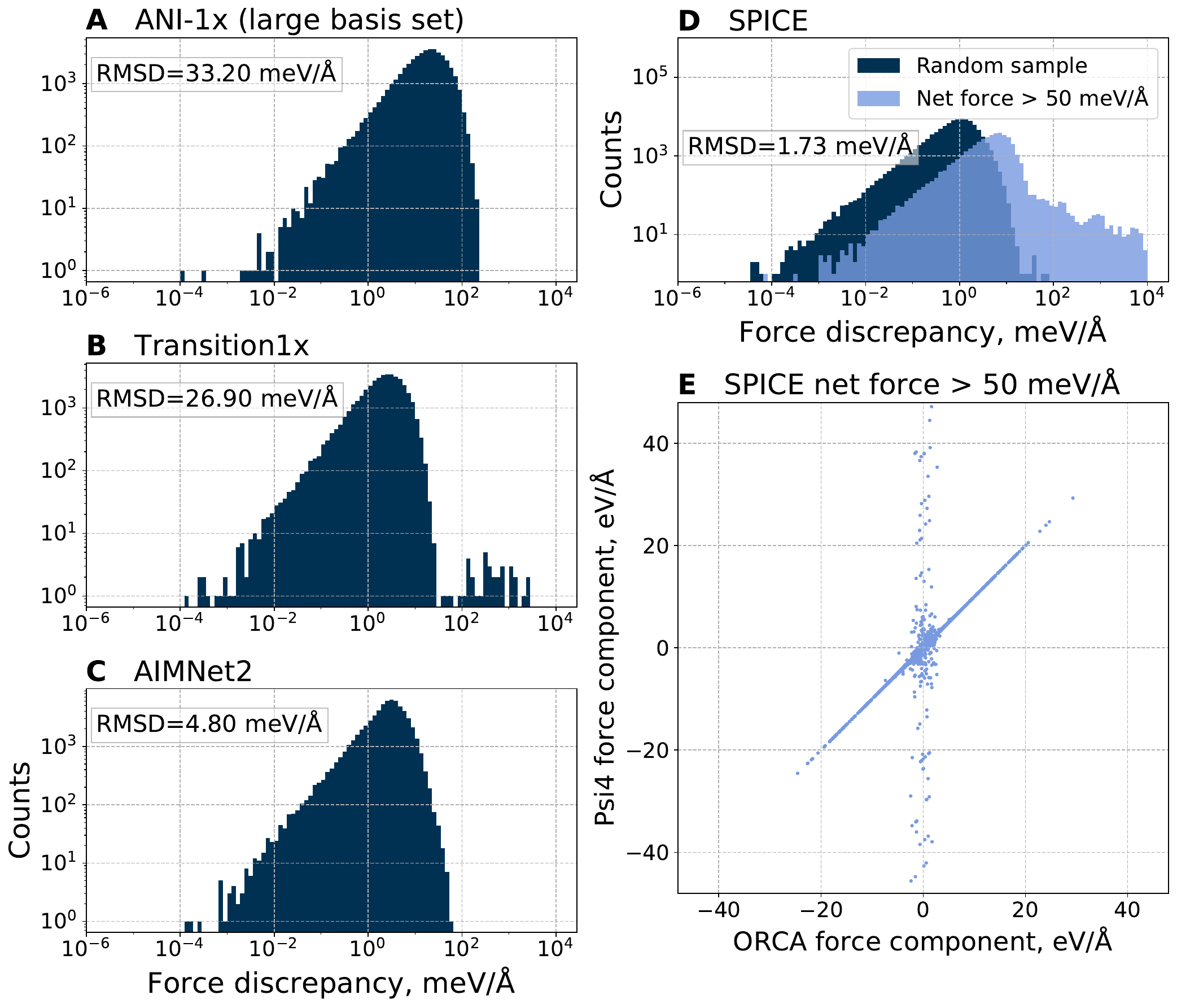}
    \caption{Absolute discrepancies in atomic Cartesian force components between original literature values and values recomputed using ORCA~6.0.1 with our chosen settings in randomly selected structures from the ANI-1x (large basis set data), Transition1x, AIMNet2, and SPICE datasets, with their force root mean square deviation (RMSD) values, shown in panels \textbf{A}, \textbf{B}, \textbf{C}, and \textbf{D}, respectively. Each data point represents the difference in a single Cartesian force component (x, y, or z) for one atom in meV/Å.
    The histograms depict the distribution of these per-component discrepancies, and the RMSD values are calculated with the component-wise discrepancies. For the SPICE dataset in panel \textbf{D}, we also provide force discrepancies for all configurations whose original Psi4 net forces are greater than 50 meV/\AA. Panel \textbf{E} compares original Psi4 and recomputed ORCA force components. Note that the scale in panel \textbf{E} has data in units of eV/\AA, while other panels report data in meV/\AA.}
    \label{fig:force_diff}
\end{figure}

\begin{figure*}[ht!]
\includegraphics[width=\textwidth]{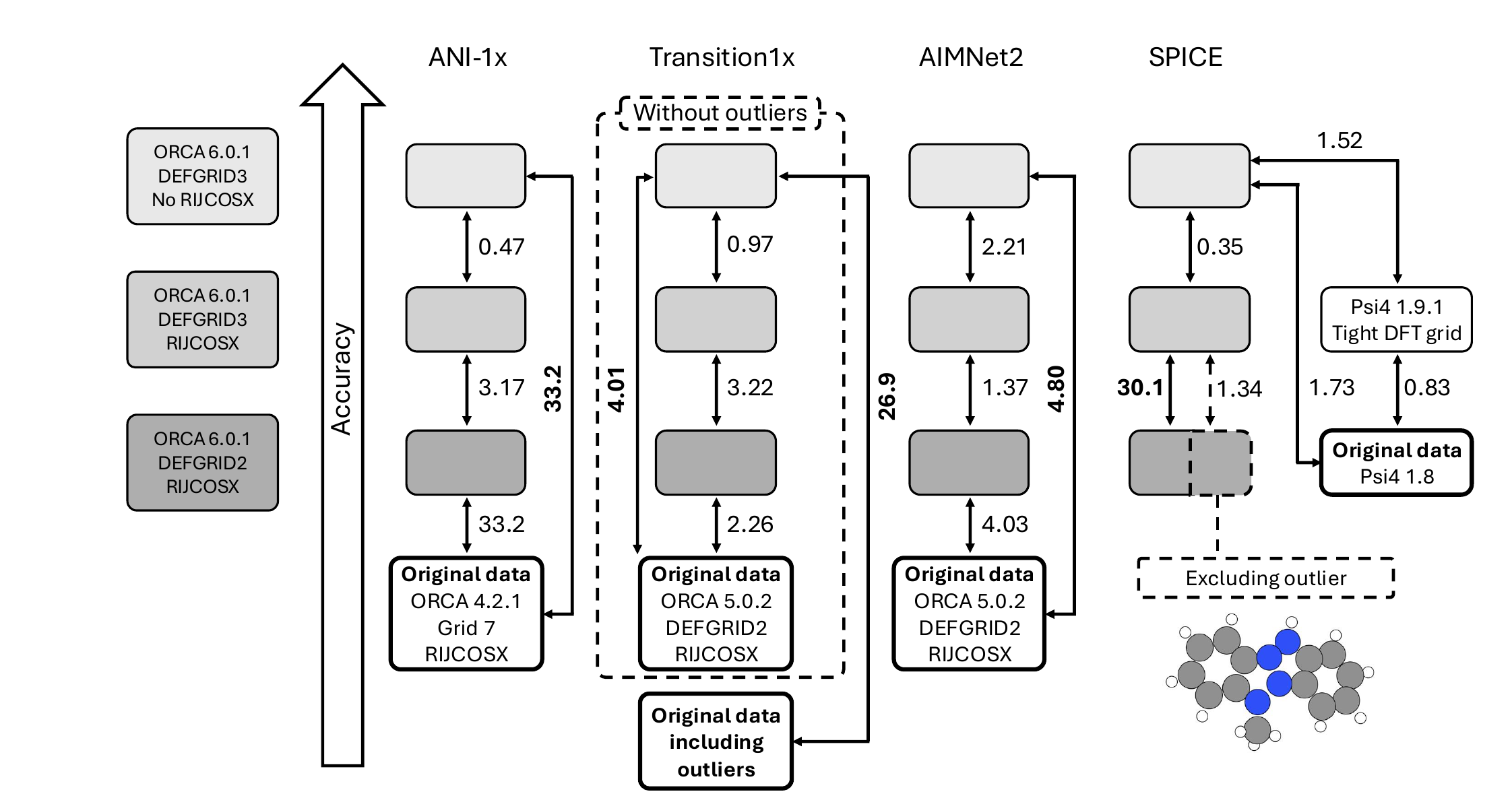}
    \caption{Force RMSDs, in meV/\AA, on the random samples containing 1000 configurations each shown in \autoref{fig:force_diff}. Each RMSD value is shown next to an arrow connecting two boxes, and the boxes represent computational settings either shown on the left hand-side in the same shade of grey, or written explicitly in the box. For Transition1x, we also show the effect of retaining outliers on the force RMSD. For SPICE, we show the single outlier which alone is responsible for changing the RMSD from 1.34 to 30.1~meV/\AA.}
    \label{fig:force_rmsds}
\end{figure*}

\autoref{fig:force_diff} summarizes the absolute discrepancies between literature and recomputed Cartesian force components for each atom individually. 
We show distributions of force component discrepancies for a sample of 1000 configurations for each dataset.
The ANI-1x (large basis set data) and Transition1x dataset samples in panels \textbf{A} and \textbf{B} show large RMSDs of 33.2 and 26.9~meV/\AA, respectively. Force discrepancy distributions show that in ANI-1x, the bulk of the data have large force discrepancies, while in Transition1x the large force RMSD is due to a small number of outliers. Exclusion of these outliers brings the RMSD down to 4.01~meV/\AA. 
The AIMNet2 and SPICE samples in panels \textbf{C} and \textbf{D} have force RMSD values of 4.80~meV/\AA\ and 1.73~meV/\AA, respectively, an order of magnitude lower than the other datasets. The SPICE sample has configurations from individual subsets it is composed of, which we show in detail in section S2 of the supplementary material.

Species with large net forces tend to also exhibit large force errors. For illustration, we use the SPICE dataset. Since we do not obtain a representative sample of outliers due to the small sample size, we also show force discrepancies in all SPICE configurations with original Psi4 net forces greater than 50~meV/\AA. As shown in panel \textbf{D}, configurations with the largest net forces also have generally greater force discrepancies compared to configurations from a random sample. Comparison of original Psi4 and recomputed ORCA force components in panel \textbf{E} shows two sets of data that follow different trends. One set is characterized by force disagreements below 10~meV/\AA, which are small relative to the magnitude of the force component. The second set is characterized by force discrepancies as large as tens of thousands of meV/\AA. Surprisingly, these large discrepancies occur only in force components with values smaller than 3000~meV/\AA\ as obtained with our converged DFT settings. We find that, within a given structure, large force discrepancies of \textgreater 50 meV/\AA\ are often associated with the presence of bromine or iodine in the structure, as shown in Figure S3 of the supplementary material.

In \autoref{fig:force_rmsds}, we show force discrepancies that arise as we vary one computational setting at a time. For each sample with 1000 configurations, we calculate the RMSD in forces between different computational settings, including the DFT grid size, the use of RIJCOSX approximation, and code version. We investigate the effect of switching on the RIJCOSX approximation while using tight DFT grids provided by the \texttt{DEFGRID3} keyword in ORCA 6.0.1. Switching on this approximation results in RMSD values below 1~meV/\AA\ for the ANI-1x, Transition1x and SPICE samples and 2.21~meV/\AA\ for the AIMNet2 sample. Such errors from the RIJCOSX approximation are small on the scale of an expected MLIP force RMSE. Next, we consider how forces are affected by reducing the DFT grid sizes to the defaults provided by the \texttt{DEFGRID2} keyword. The forces change with RMSD  values of 1-3~meV/\AA\ in the ANI-1x, Transition1x, and AIMNet2 samples and by 30.1~meV/\AA\ in the SPICE sample. The RMSD of the SPICE sample is affected by one outlier structure with force discrepancies of several eV/\AA, which is why we report RMSD values with and without the outlier. Based on the ANI-1x and Transition1x sample RMSD values and the outlier in the SPICE sample, we suggest that the \texttt{DEFGRID3} tight grid settings are required for reliable forces. We also compare ORCA 6.0.1 forces to forces from older ORCA versions with less accurate DFT grids which were used to produce the original datasets. In particular, changing to ORCA 4.2.1 for the ANI-1x sample results in an RMSD of 33.2~meV/\AA. For the sample of SPICE, which was originally computed using Psi4 1.8.1 or 1.8.2, we recomputed the forces with Psi4 1.9.1 using tighter DFT grid settings to reduce nonzero net forces. The force RMSD between the forces recomputed with converged ORCA settings and recomputed Psi4 forces is 1.52~meV/\AA, close to the 1~meV/\AA\ threshold, showing consistency between the different programs if converged settings are used.

\begin{figure}[h!]
\includegraphics[width=\columnwidth]{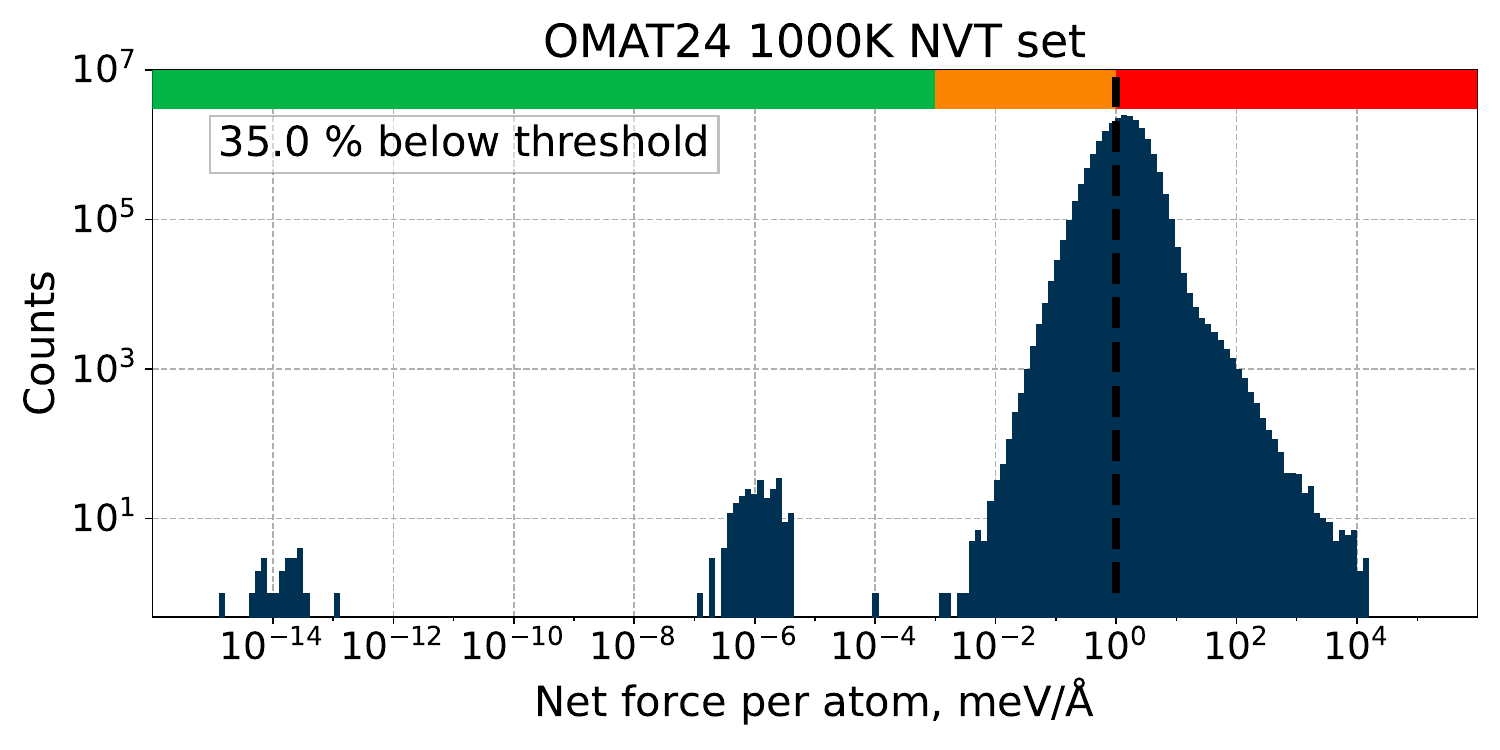}
    \caption{Distribution of per atom net forces in the 1000 K NVT subset of OMAT24. 65\%\ of the configurations exceed 1~meV/\AA/atom net force threshold, with thousands even exceeding 100~meV/\AA/atom. The red bar on top indicates the region above the 1~meV/\AA/atom threshold. The green bar indicates negligible net forces below $10^{-3}$~meV/\AA/atom and the amber bar indicates net forces of intermediate size.}
    \label{fig:omat}
\end{figure}

Although we focus on molecular datasets, net forces are relevant for periodic systems, too. For example, we find negligible net forces of less than $10^{-5}$~meV/\AA/atom in a 10\%\ subset of the Materials Project Trajectory (MPTrj) \cite{deng_2023_chgnet} dataset. In addition, we find significant nonzero DFT net forces in the 1000 K NVT subset from the OMAT24 \cite{omat24} materials dataset, with thousands of configurations exhibiting net forces exceeding 100~meV/\AA/atom, as shown in \autoref{fig:omat}. We find that the nonzero force problem is solved by using a tighter energy convergence threshold specified by the \texttt{EDIFF} keyword in VASP\cite{PhysRevB.47.558}, concluding that the threshold of $5\times 10^{-5}$ eV per atom used in OMAT24 is too loose. An example structure from OMAT24 computed with different energy convergence thresholds is given in section S4 of the supplementary material. In the same section, we also collect net forces for other periodic systems computed with VASP and other popular periodic DFT codes including CASTEP \cite{castep}, CP2K \cite{cp2k}, and FHI-aims \cite{fhi-aims}.

To reiterate, DFT forces are included in molecular datasets with the primary goal to provide training data for MLIPs. Therefore, we draw the reader's attention to the potential effect of the demonstrated force discrepancies on the quality of an MLIP trained using these data. We may consider force discrepancies due to numerical precision of DFT as a contribution to the overall MLIP error. In principle, an MLIP may be able to cope with noisy DFT training data. However, it is difficult to assess the accuracy of the MLIP using test data that are noisy as well. Therefore, errors in DFT forces should ideally be much smaller than the expected MLIP force errors. Since general-purpose MLIP force errors can be as low as 10~meV/\AA, DFT force RMSD of 4.8~meV/\AA, as in AIMNet2, is of the size where DFT force discrepancies start to affect the accuracy of the MLIP. On the other hand, force RMSDs of around 30~meV/\AA, as in ANI-1x, and Transition1x without the removal of outliers, will significantly limit the overall accuracy of the models trained on these data.

To conclude, large DFT datasets are incredibly valuable for the modeling of molecules and materials.
However, we have shown that the DFT data of numerous molecular datasets suffer from unphysical nonzero net forces, which are a symptom of numerical problems in the DFT calculation. At times, the force discrepancies are significant, reaching or exceeding the force errors expected from state-of-the-art general-purpose MLIPs. We have established settings that allow us to obtain reliable DFT reference forces and used them to quantify the error in the DFT forces by computing reliable reference forces for fractions of those datasets.  
These results lead to important guidelines for considering existing datasets and the development of new datasets. For existing datasets, care must be taken regarding the quality of the DFT data. Low quality DFT forces can be addressed to some extent by filtering. For the development of new datasets, good quality DFT settings should be chosen to minimize numerical errors and the associated noise. Net forces are an indicator of force errors, and we recommend that more attention is paid to them. As new datasets of molecules and materials are being developed, and MLIP architectures are able to fit reference data with increasing accuracy, we emphasize that labelling databases with high-quality DFT data is increasingly important.

\section*{Supplementary material}

See the supplementary material for computational details, overview of net forces and force discrepancies in the SPICE dataset subsets, examples of molecules in the SPICE dataset with large DFT force errors, and additional examples of net forces data from periodic DFT calculations.

\begin{acknowledgments}

D.K. acknowledges support from the EPSRC Centre for Doctoral Training in Automated Chemical Synthesis Enabled by Digital Molecular Technologies (SynTech) with grant reference EP/S024220/1.
A.M. acknowledges support from the European Union under the “n-AQUA” European Research
Council project (Grant No. 101071937). F.B. acknowledges support from the Alexander von Humboldt Foundation through a Feodor Lynen Research Fellowship, from the Isaac Newton Trust through an Early Career Fellowship, and from Churchill College, Cambridge, through a Postdoctoral By-Fellowship.
This work was performed using resources provided by the HEC Materials Chemistry Consortium funded by the Engineering and Physical Sciences Research Council (grant number EP/R029431 and EP/X035859) and the Cambridge Service for Data Driven Discovery (CSD3) operated by the University of Cambridge Research Computing Service (www.csd3.cam.ac.uk), provided by Dell EMC and Intel using Tier-2 funding from the Engineering and Physical Sciences Research Council (capital grant EP/T022159/1), and DiRAC funding from the Science and Technology Facilities Council (www.dirac.ac.uk).
We thank Kaifeng Niu, Samuel Coles, Xavier R. Advincula, Eszter Varga-Umbrich, and Giaan Kler-Young for providing structures and calculations for the supplementary material.

\end{acknowledgments}

\section*{Author declarations}

\subsection*{Author contributions}

Domantas Kuryla: Conceptualization (equal); Data curation (equal); Formal analysis (equal); Investigation (equal); Validation (equal); Visualization (equal); Writing – original draft (equal).

Fabian Berger: Conceptualization (equal); Formal analysis (equal); Supervision (equal); Writing – review and editing (equal).

G\'abor Cs\'anyi: Conceptualization (equal); Formal analysis (equal); Project administration (equal); Resources (equal); Supervision (equal); Writing – review and editing (equal).

Angelos Michaelides: Conceptualization (equal); Formal analysis (equal); Funding acquisition (equal); Project administration (equal); Resources (equal); Supervision (equal); Writing – review and editing (equal).

\subsection*{Conflict of interest}

G. C. has equity interest in Symmetric Group LLP that licenses force fields commercially, and in Ångstrom AI.

\subsection*{Data availability}

Computed net forces for all datasets discussed in this work and recomputed DFT forces for random data samples are available on GitHub at \href{https://github.com/water-ice-group/datasets_dft_accuracy}{\texttt{water-ice-group/datasets\_dft\_accuracy}}.

\renewcommand{\bibsection}{\section*{References}}

\bibliography{bibliography}

% \input{SI}
% \appendix
%\clearpage
%\begin{widetext}
%\input{SI_arxiv}
%\end{widetext}

%\includepdf[pages=-,pagecommand={},width=\textwidth]{supplementary_material.pdf}
\end{document}

\preprint{AIP/123-QED}

\title{Supplementary material for ``How Accurate Are DFT Forces? Unexpectedly Large Uncertainties in Molecular Datasets''}

\author{Domantas Kuryla}
\affiliation{Yusuf Hamied Department of Chemistry, University of Cambridge, Lensfield Road, Cambridge, UK}
\affiliation{Engineering Laboratory, University of Cambridge, Trumpington St and JJ Thomson Ave, Cambridge, UK}

\author{Fabian Berger}
\email{fb593@cam.ac.uk}
\affiliation{Yusuf Hamied Department of Chemistry, University of Cambridge, Lensfield Road, Cambridge, UK}

\author{G\'abor Cs\'anyi}
\email{gc121@cam.ac.uk}
\affiliation{Engineering Laboratory, University of Cambridge, Trumpington St and JJ Thomson Ave, Cambridge, UK}

\author{Angelos Michaelides}
\email{am452@cam.ac.uk}
\affiliation{Yusuf Hamied Department of Chemistry, University of Cambridge, Lensfield Road, Cambridge, UK}

\setlength{\parindent}{0cm}

\doublespacing
\maketitle

In the supplementary material we provide:

\begin{itemize}
    \item[\ref{computational}] Computational settings: ORCA and Psi4 computational setups used in this work.
    \item[\ref{subsets}] Net forces and force dicrepancies in the susbets of the SPICE dataset.
    \item[\ref{spice_outliers}] Examples of stuctures in the SPICE dataset with particularly large Psi4 net forces.
    \item[\ref{periodic_codes}] Net forces obtained with periodic DFT codes in various periodic systems
\end{itemize}

\clearpage

\renewcommand{\thesection}{S\arabic{section}}
\renewcommand{\thetable}{S\arabic{table}}
\renewcommand{\thefigure}{S\arabic{figure}}

\section{Computational settings}\label{computational}

The random samples from ANI-1x \cite{smith_ani-1ccx_2020}, Transition1x \cite{schreiner_transition1x_2022}, AIMNet2 \cite{anstine_aimnet2_2025}, and SPICE \cite{eastman_spice_2023, eastman_nutmeg_2024} datasets discussed in Figures 2 and 3 of the main text are recomputed using ORCA 6.0.1 \cite{neese_software_2025} at $\omega$B97x/def2-TZVPP, $\omega$B97x/6-31G(d), $\omega$B97M-D3(BJ)/def2-TZVPP, and $\omega$B97M-D3(BJ)/def2-TZVPPD levels of theory, respectively. In all calculations, we use self-consistent field (SCF) convergence criteria provided by the \texttt{TightSCF} keyword, which include a $10^{-8}$~Hartree threshold for energy change between cycles. For the most accurate ORCA settings, we disable the RIJCOSX \cite{rijcosx} approximation and use tight DFT grid settings provided by the \texttt{DEFGRID3} keyword. For comparison to more approximate DFT settings, as described in the main text, we use the RIJCOSX approximation and default DFT grid settings provided by the \texttt{DEFGRID2} keyword. To reproduce the original ANI-1x $\omega$B97x/def2-TZVPP data, we use ORCA 4.2.1 with the tightest DFT grids provided by the \texttt{Grid 7} keyword. For Transition1x specifically, we use the unrestricted Kohn-Sham (UKS) formalism, as appropriate for reactive configurations. 

To obtain more accurate SPICE dataset \cite{eastman_spice_2023} forces, we use Psi4 1.9.1 \cite{smith_p_2020}. To increase accuracy and reduce net forces, we use tighter than default DFT integration grids, with 99 radial points and 1202 spherical points, provided using the \texttt{dft\_radial\_points} and \texttt{dft\_spherical\_points} keywords, respectively.

\clearpage

\section{Net forces and force discrepancies in SPICE subsets}\label{subsets}

\begin{figure}[h!]
\includegraphics[width=\columnwidth]{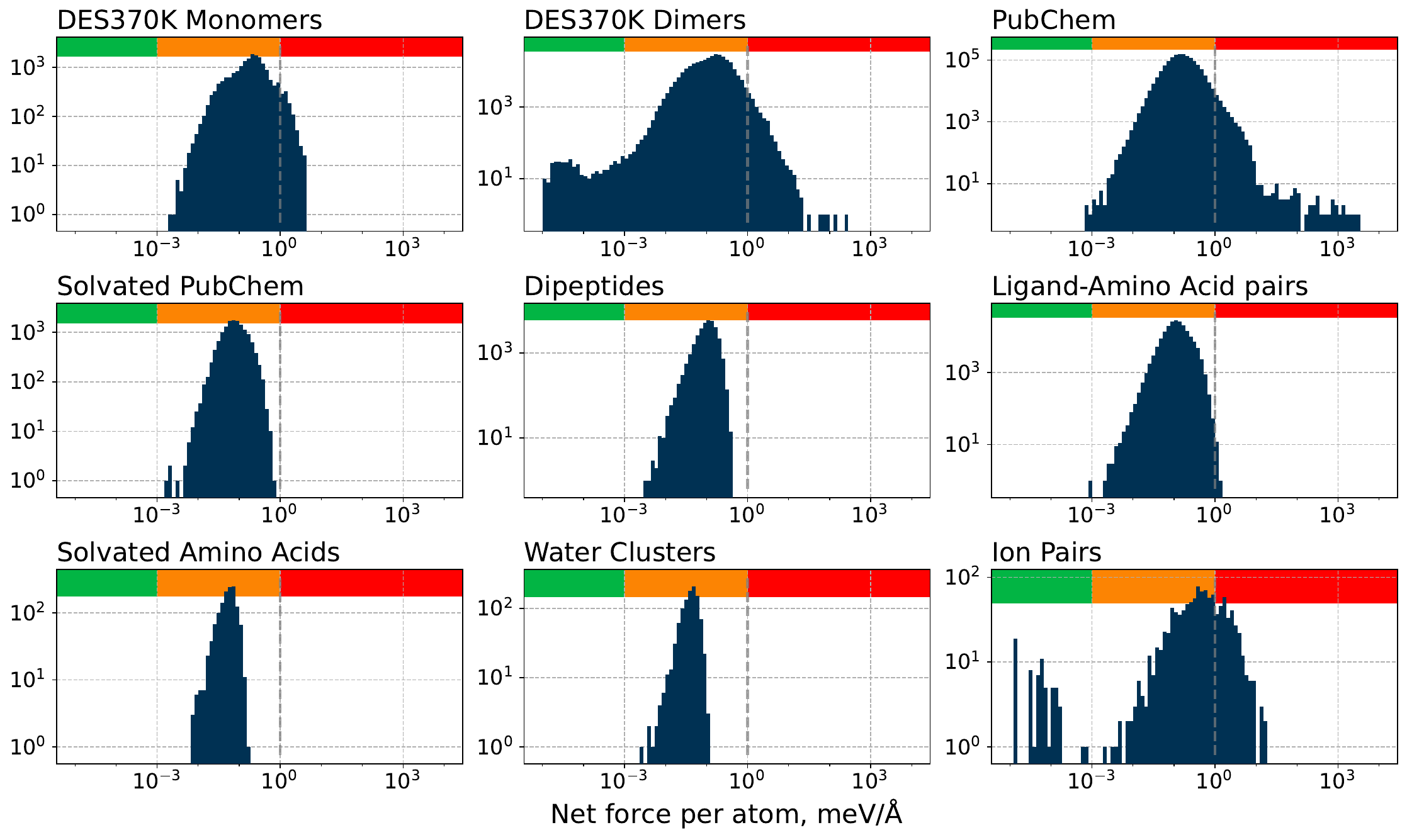}
    \caption{Per atom net force distributions in subsets of the SPICE dataset. Configurations with the largest net forces belong to the PubChem drug-like molecule subset. In addition, the DES370K Monomers and Dimers subsets, and the Ion Pairs subset have configurations with net forces exceeding 1 meV/\AA. The red bar at the top of each plot highlights net forces above 1~meV/\AA/atom which indicate significant errors in individual force components.  In the region between $10^{-3}$ and 1~meV/\AA/atom, indicated with an amber bar, significant DFT force errors often appear but do not result in large net forces. Negligible net forces below $10^{-3}$~meV/\AA/atom are in the region indicated with the green bar.}
    \label{fig:spice_net_forces}
\end{figure}

\begin{figure}[h!]
\includegraphics[width=12cm]{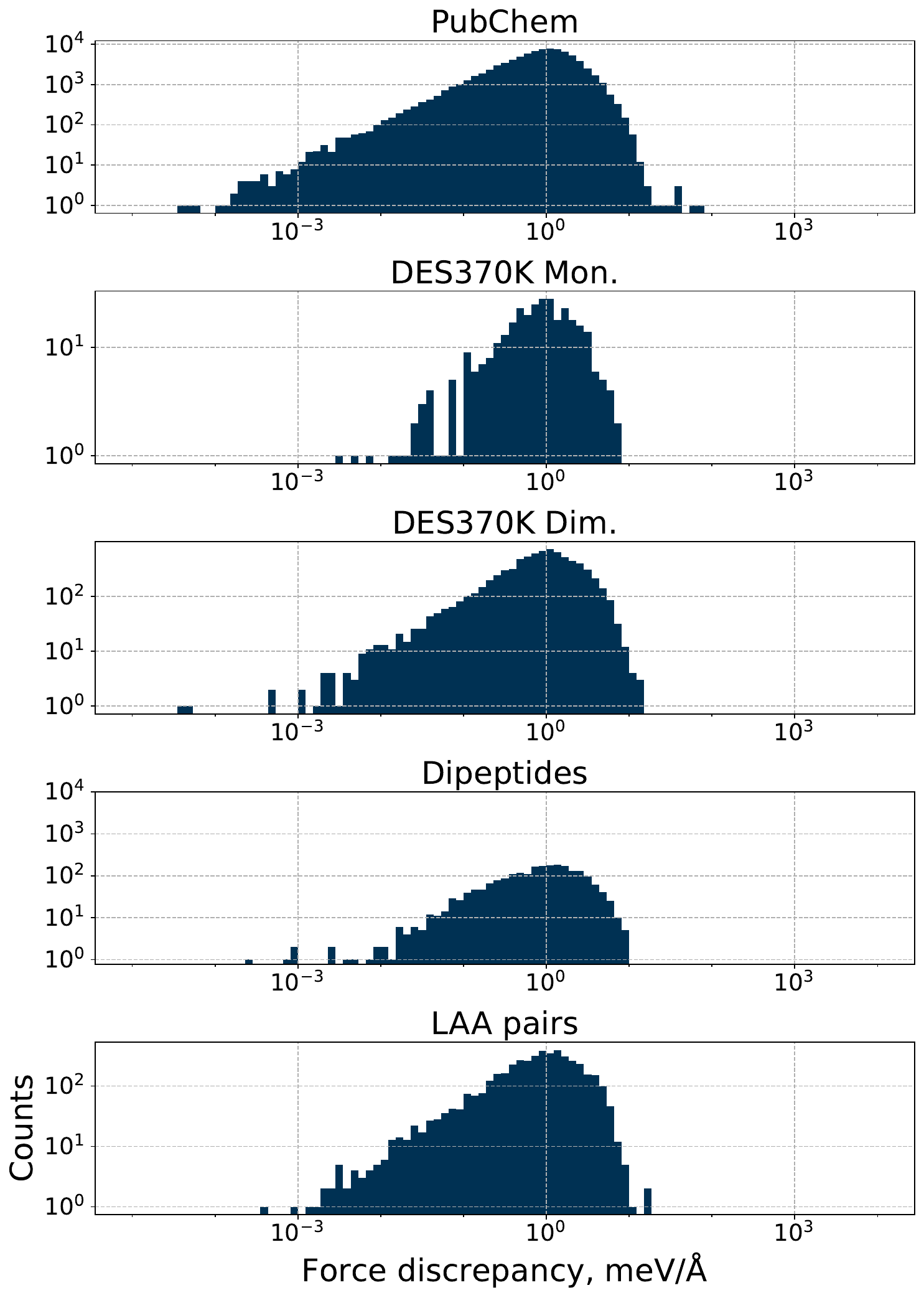}
    \caption{Discrepancies in force components in the 1000 configuration SPICE sample, broken down by subset. Out of 9 SPICE subsets, only 5 contribute to this sample, since some subsets are very small. All 5 subsets follow similar distributions, with force discrepancies peaking at 1~meV/\AA.}
    \label{fig:subsets_force_diff}
\end{figure}

In the main text, we show the distributions of per atom net forces in various datasets. The SPICE dataset consists of various subsets with different distributions of net forces. In \autoref{fig:spice_net_forces}, we show these distributions. In addition, in \autoref{fig:subsets_force_diff}, we show discrepancies between reported and recomputed forces for the 1000 configuration sample from SPICE, broken down by subset.

\clearpage

\section{Examples of large net force structures in the SPICE dataset}\label{spice_outliers}

In this section, we show a selection of molecules from the SPICE dataset with particularly large net forces. We also indicate the largest force discrepancies on individual atoms. The force discrepancies shown here are calculated by subtracting the original Psi4 force vector from the recomputed ORCA force vector and taking the norm. Generally, the very large force discrepancies are associated with only a few atoms in each structure. 

\begin{figure}[h!]
\includegraphics[width=\textwidth]{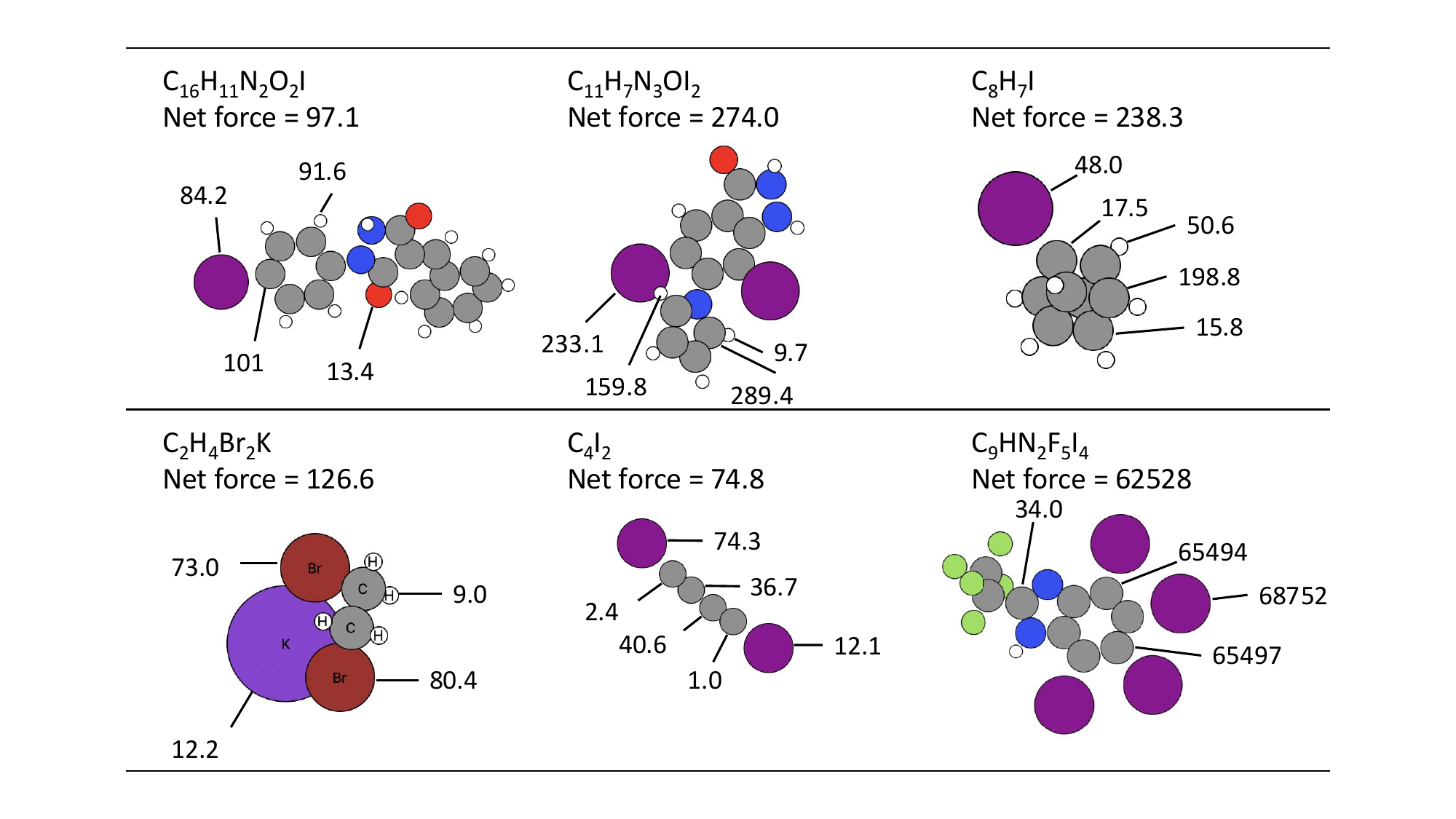}
    \caption{Examples of SPICE molecules among those with the largest original Psi4 net forces. For each given molecule, several atoms with the largest discrepancies in original Psi4 - recomputed ORCA forces are indicated with the norm of the force difference vector in meV/\AA.}
    \label{fig:spice}
\end{figure}

\clearpage

\section{Net forces in various periodic systems}\label{periodic_codes}

In this section, we provide the net forces calculated for various condensed phase systems using periodic DFT codes including VASP \cite{PhysRevB.47.558}, CP2K \cite{cp2k}, CASTEP \cite{castep}, and FHI-aims\cite{fhi-aims}. In particular, we show that large net forces in VASP can result from insufficiently tight convergence of the Kohn-Sham orbitals. In addition, CP2K calculations give relatively large net forces on the order of 10 to 1000~meV/\AA\ in systems of 100 to 500 atoms.

\subsection{Systems computed using VASP}

In \autoref{table:vasp_omat}, we show how the net force is affected by the energy convergence threshold in a periodic \ch{CS3F} system from the OMAT24 \cite{omat24} dataset. The calculations are done using the PBE \cite{pbe} functional with a plane wave cutoff of 520 eV. Net forces are computed with a loose energy convergence threshold of $5\times 10^{-5}$ eV per atom, following the OMAT24 guidelines, specified using the \texttt{EDIFF} keyword, and a tight energy convergence threshold of $10^{-7}$ eV. For this system, we use a 7×7×7 k-point grid. The input settings are generated using the pymatgen package \cite{pymatgen} as it was used in the development of OMAT24.
 
\begin{table}[h!]
\normalsize
\centering
\begin{tabular}{p{5cm} p{6cm} p{4cm}}
\toprule
\textbf{EDIFF, eV} & \textbf{Net force vector (XYZ), meV/\AA} & \textbf{Net force magnitude, meV/\AA} \\
\midrule
(a) 2.5×$10^{-4}$ & -1.446, 12.581, 7.818 & 14.883 \\
(b) $10^{-7}$ & -0.178, 0.036, -0.0071 & 0.182 \\
\bottomrule
\end{tabular}
\caption{Net forces on a periodic \ch{CS3F} system computed with VASP with different energy difference tolerances.}
\label{table:vasp_omat}
\end{table}

The following systems in \autoref{table:vasp-surfaces} are computed using the PBE functional with the DFT-D3 dispersion correction \cite{grimme_consistent_2010}. A 400 eV plane wave cutoff is used. The k-point sampling is done on a 2×2×1 grid centered on the $\Gamma$-point. The calculations use a $10^{-6}$ eV energy convergence threshold.
%and a 0.01 eV/\AA\ gradient convergence threshold.

Net forces are small but highly system-dependent: the net force differs by 9 orders of magnitude for the same Pt surface with different numbers of adsorbed H atoms. In addition, net forces in system (b) are much greater in the x, y directions than in the z direction.

\begin{table}[h!]
\centering
\normalsize
%\begin{adjustbox}{width=\textwidth}
\begin{tabular}{p{5cm} p{6cm} p{4cm}}
\toprule
\textbf{System} & \textbf{Net force vector (XYZ), meV/\AA} & \textbf{Net force magnitude, meV/\AA} \\
\midrule
(a) 64 H atom layer on Pt(111)& 0.00, -2.65$\times10^{-12}$,  7.13$\times10^{-12}$ & 7.61$\times10^{-12}$ \\
(b) 32 H atom island on Pt(111) & 3.00$\times10^{-3}$, 2.00$\times10^{-3}$, 2.57$\times10^{-12}$ & 3.61$\times10^{-3}$ \\
(c) H atom layer on Cu(111) & 0.00, -6.25$\times10^{-13}$, 1.96$\times10^{-13}$ & 6.55$\times10^{-13}$ \\
\bottomrule
\end{tabular}
%\end{adjustbox}
\caption{Net forces of selected surface systems computed with VASP. Net forces are negligible, with system (b) having very different net forces in different directions, on the order of $10^{-3}$ meV/\AA\ in the x, y directions and on the order of $10^{-12}$ meV/\AA\ in the z direction.}
\label{table:vasp-surfaces}
\end{table}

\clearpage

\subsection{Systems computed using CP2K}

The following is a bulk water system of 64 water molecules. The single point calculations are carried out using the revPBE \cite{revpbe} functional and TZV2P basis sets. An energy convergence threshold of $10^{-8}$ Ry is used in the inner SCF loop. One setting that can control the convergence of the electron density and the quality of the forces is the plane wave cutoff applied to the finest level grid specified by the \texttt{CUTOFF} keyword. In \autoref{table:cp2k-water}, we show that net forces can be reduced by using a tighter cutoff. Specifically, using a cutoff of 1200 Ry resulted in net force 6 times smaller than when cutoffs of 300 and 600 Ry were used.

\begin{table}[h!]
\normalsize
\centering
\begin{tabular}{p{5cm} p{6cm} p{4cm}}
\toprule
\textbf{Grid cutoff, Ry} & \textbf{Net force vector (XYZ), meV/\AA} & \textbf{Net force magnitude, meV/\AA} \\
\midrule
300 & -167.04, -134.87, 88.21 & 232.10 \\
600 & 190.03, -97.42, 118.86 & 244.40 \\
1200 & 22.03, -32.17,  12.43 & 40.92 \\
%Bulk water (64 molecules) & 58.88, 25.92, 27.97 & 70.15 \\
%Water monolayer in vacuum (144 \ch{H2O}) & 1.080, -53.79, -40.21 & 67.17 \\
%\makecell{Water monolayer nanoconfined\\between graphene sheets (120\ch{C}, 15 \ch{H2O})} & 3.959, 95.44,  7.148 & 95.79\\
\bottomrule
\end{tabular}
\caption{Net forces in a 64 water box with different grid cutoffs. Net forces are reduced by using a tighter grid cutoff of 1200 Ry but are still too large to be considered negligible, at tens of meV/\AA.}
\label{table:cp2k-water}
\end{table}

\begin{figure}[h!]
\includegraphics[width=12cm]{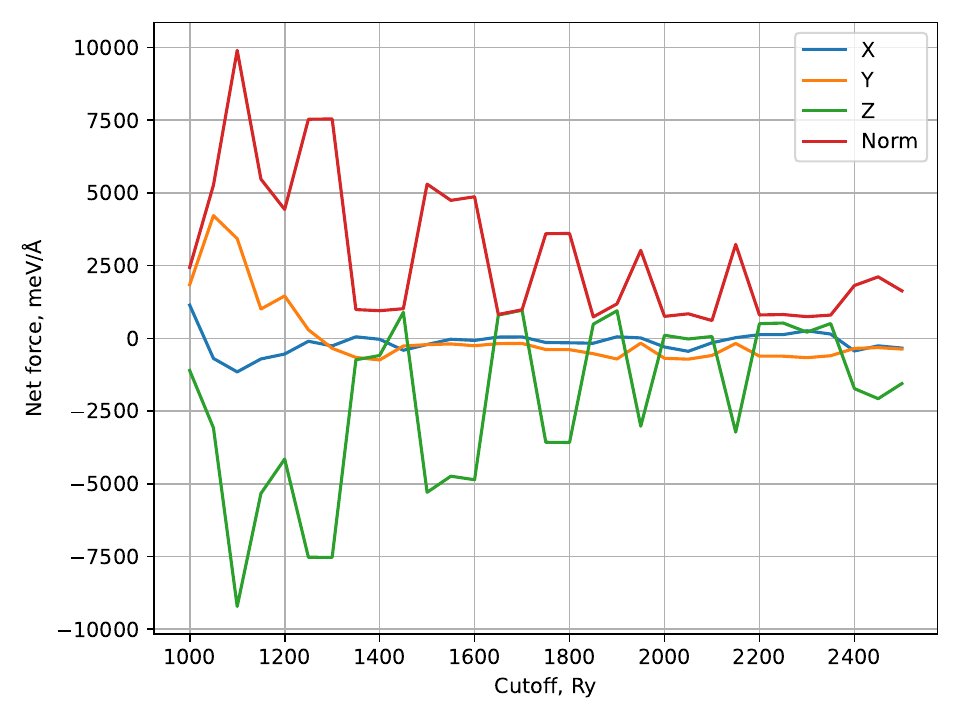}
    \caption{Norm of the net force and Cartesian components of the net force for a MgO-water interface computed with different plane wave cutoff values. While in the X and Y directions, parallel to the interface plane, the net force goes down to several hundred meV/\AA\ for large cutoff values, in the Z direction perpendicular to the interface the net force fluctuates from several hundred to thousands of meV/\AA.}
    \label{fig:cp2k-mgo-water}
\end{figure}

For interfaces, we find that obtaining converged forces in CP2K can be more challenging than for homogeneous systems such as bulk water. The following example is an MgO-water interface computed using the opt88-vdW \cite{opt88} functional and the \texttt{GTH-PBE} pseudopotential. For hydrogen and oxygen, the \texttt{DZVP-MOLOPT-GTH} basis set is used, and \texttt{DZVP-MOLOPT-SR-GTH} for magnesium. The electron density is converged with a total energy tolerance of $10^{-6}$ Ry. In \autoref{fig:cp2k-mgo-water} we show the variation in net force and Cartesian net force components as the plane wave cutoff is varied from 1000 to 2500 Ry for a relative cutoff, specified by the \texttt{REL\_CUTOFF} keyword, of 40 Ry. For low values of the cutoff of 1000-1200 Ry, the net force is on the order of several thousand meV/\AA. For large values of the cutoff, the X and Y components are reduced to several hundred meV/\AA, while the Z component (perpendicular to the interface) fluctuates from several hundred to thousands of meV/\AA.

\clearpage

\subsection{Systems computed using FHI-aims}

In \autoref{table:fhi-aims}, we provide net forces computed for magnesium oxide using FHI-aims. The calculation was carried out using the revPBE functional with the DFT-D3 dispersion correction, using the 'tight' level of settings for basis sets and integration grids provided in FHI-aims, and a 2×2×1 k-point grid. These settings provide well converged results with negligible net forces. 

\begin{table}[h!]
\normalsize
\centering
\begin{tabular}{p{5cm} p{6cm} p{4cm}}
\toprule
\textbf{System} & \textbf{Net force vector (XYZ), meV/\AA} & \textbf{Net force magnitude, meV/\AA} \\
\midrule
Bulk MgO, 8-atom unit cell & -2.00$\times10^{-8}$,  1.17$\times10^{-8}$, -1.72$\times10^{-8}$ & 2.89$\times10^{-8}$ \\
\bottomrule
\end{tabular}
\caption{Net forces in a bulk MgO system computed with FHI-aims.}
\label{table:fhi-aims}
\end{table}

\subsection{Systems computed using CASTEP}

In \autoref{table:castep}, we provide net forces computed for iron and iron-hydrogen systems using CASTEP. The systems are computed using the PBE functional with a 4×4×1 k-point grid and a 500 eV plane wave cutoff. The electron density is converged with an energy convergence threshold of $10^{-7}$ eV. The net forces are generally on the order of 0.01 meV/\AA, much greater than the net forces observed with the example computed with the FHI-aims code, but still small enough that the errors causing these net forces are negligible compared to MLIP force errors.

\begin{table}[h!]
\normalsize
\centering
\begin{tabular}{p{5cm} p{7cm} p{3cm}}
\toprule
\textbf{System} & \textbf{Net force vector (XYZ), meV/\AA} & \textbf{Net force magnitude, meV/\AA} \\
\midrule
Bulk BCC iron & $-5.00\times10^{-2}$, $1.00\times10^{-2}$, $1.36\times10^{-17}$ & 5.10$\times10^{-2}$ \\
Iron slab with interstitial H atoms & $-1.00\times10^{-2}$, $-3.00\times10^{-2}$, $-5.00\times10^{-2}$ & $5.91\times10^{-2}$ \\
Iron slab with H atom monolayer & $-1.00\times10^{-2}$,  $1.00\times10^{-2}$, $-8.33\times10^{-13}$ & 1.41$\times10^{-2}$ \\
\bottomrule
\end{tabular}
\caption{Net forces in selected iron-hydrogen systems computed with CASTEP.}
\label{table:castep}
\end{table}

\setcounter{section}{0}

\clearpage

\renewcommand{\bibsection}{\section*{References}}

\bibliography{bibliography}